\documentclass{ws-mpla}
\usepackage[super]{cite}
\usepackage{graphicx}
\usepackage{braket}
\usepackage{bm}
\usepackage{comment}
\usepackage{here}
\usepackage{color}
\begin{document}
\def\Journal#1#2#3#4{{#1} {\bf #2}, #3 (#4)}
\def\AHEP{Advances in High Energy Physics.} 	
\def\AACA{Advances in Applied Clifford Algebras} 	
\def\ARNPS{Annu. Rev. Nucl. Part. Sci.} 
\def\AandA{Astron. Astrophys.} 
\def\ANP{Ann. Phys.}
\def\APJ{Astrophys. J.}
\def\APJS{Astrophys. J. Suppl}
\def\COMR{Comptes Rendues}
\def\CQG{Class. Quantum Grav.}
\def\CPC{Chin. Phys. C}
\def\EPJC{Eur. Phys. J. C}
\def\EPL{EPL}
\def\IJMPA{Int. J. Mod. Phys. A}
\def\IJMPE{Int. J. Mod. Phys. E}
\def\JCAP{J. Cosmol. Astropart. Phys.}
\def\JHEP{J. High Energy Phys.}
\def\JETPL{JETP. Lett.}
\def\JETPUSSR{JETP (USSR)}
\def\JPG{J. Phys. G} 
\def\JPCS{J. Phys. Conf. Ser.} 
\def\JPGNP{J. Phys. G: Nucl. Part. Phys.} 
\def\JNP{J. of Number Theor.} 
\def\MPLA{Mod. Phys. Lett. A}
\def\NIMA{Nucl. Instrum. Meth. A.}
\def\NATU{Nature}
\def\NCA{Nuovo Cimento}
\def\NJP{New. J. Phys.}
\def\NPB{Nucl. Phys. B}
\def\NPBOLD{Nucl. Phys.}
\def\NPBSUPPL{Nucl. Phys. B. Proc. Suppl.}
\def\PL{Phys. Lett.}
\def\PLB{{Phys. Lett.} B}
\def\PMCA{PMC Phys. A}
\def\PREP{Phys. Rep.}
\def\PPNP{Prog. Part. Nucl. Phys.}
\def\PLBOLD{Phys. Lett.}
\def\PAN{Phys. Atom. Nucl.}
\def\PRL{Phys. Rev. Lett.}
\def\PRD{Phys. Rev. D}
\def\PRC{Phys. Rev. C}
\def\PR{Phys. Rev.}
\def\PTP{Prog. Theor. Phys.}
\def\PTEP{Prog. Theor. Exp. Phys.}
\def\RMP{Rev. Mod. Phys.}
\def\RPP{Rep. Prog. Phys.}
\def\SJNP{Sov. J. Nucl. Phys.}
\def\SCIENCE{Science}
\def\SPJETP{Sov. Phys. JETP.}
\def\TNYAS{Trans. New York Acad. Sci.}
\def\ZETP{Zh. Eksp. Teor. Piz.}
\def\ZFPH{Z. fur Physik}
\def\ZPC{Z. Phys. C}
\markboth{Yuta Hyodo and Teruyuki Kitabayashi}{Unified neutrino mixing and approximate $\mu$-$\tau$ reflection symmetry}

\catchline{}{}{}{}{}

\title{Unified neutrino mixing and approximate $\mu$-$\tau$ reflection symmetry}

\author{Yuta Hyodo and Teruyuki Kitabayashi}

\address{Department of Physics, Tokai University, 4-1-1 Kitakaname, Hiratsuka, Kanagawa 259-1292, Japan\\corresponding author: yuta.h.1410@gmail.com}

\maketitle

\pub{Received (Day Month Year)}{Revised (Day Month Year)}

\begin{abstract}
We investigate the phenomenology of a unified neutrino mixing framework, which serves as the origin of well-known neutrino mixing schemes such as the tribimaximal mixing (TBM), bimaximal mixing (BM), golden ratio mixing (GRM) and hexagonal mixing (HM). Our analysis reveals that the predicted sum of neutrino masses derived from an approximate $\mu$-$\tau$ reflection symmetric flavor neutrino mass matrix based on the unified neutrino mixing with an inverted mass ordering, is excluded from DESI2024 and Supernova Ia luminosity distance data. This conclusion implies that TBM, BM, GRM, and HM, under an approximate $\mu$-$\tau$ reflection symmetry with an inverted mass ordering of neutrinos, are also excluded from observations.

\end{abstract}

\ccode{PACS Nos.: 14.60.Pq}

\section{Introduction\label{section:introduction}}
Understanding the flavor structure of neutrinos is one of the most significant challenges in particle physics. Various approaches have been explored to construct the neutrino mass matrix, such as the trimaximal texture\cite{Harrison2002PLB,Xing2002PLB,Harrison2002PLB2,Kitabayashi2007PRD}, the $\mu$-$\tau$ symmetry texture\cite{Fukuyama1997,Lam2001PLB,Ma2001PRL,Balaji2001PLB,Koide2002PRD,Kitabayashi2003PRD,Koide2004PRD,Aizawa2004PRD,Ghosal2004MPLA,Mohapatra2005PRD,Koide2005PLB,Kitabayashi2005PLB,Haba206PRD,Xing2006PLB,Ahn2006PRD,Joshipura2008EPJC,Gomez-Izquierdo2010PRD,He2001PRD,He2012PRD,Gomez-Izquierdo2017EPJC,Fukuyama2017PTEP,Kitabayashi2016IJMPA,Kitabayashi2016PRD,Bao2021arXiv,XingRPP2023,TostadoNPB2021,TostadoNPB2023,ZhaoJHEP2017,HarrisonPLB2002,XingRPP2016,TostadoarXiv2024}, zero textures\cite{Berger2001PRD,Frampton2002PLB,Xing2002PLB530,Xing2002PLB539,Kageyama2002PLB,Xing2004PRD,Grimus2004EPJC,Low2004PRD,Low2005PRD,Grimus2005JPG,Dev2007PRD,Xing2009PLB,Fritzsch2011JHEP,Kumar2011PRD,Dev2011PLB,Araki2012JHEP,Ludle2012NPB,Lashin2012PRD,Deepthi2012EPJC,Meloni2013NPB,Meloni2014PRD,Dev2014PRD,Felipe2014NPB,Ludl2014JHEP,Cebola2015PRD,Gautam2015PRD,Dev2015EPJC,Kitabayashi2016PRD1,Zhou2016CPC,Singh2016PTEP,Bora2017PRD,Barreiros2018PRD,Kitabayashi2018PRD,Barreiros2019JHEP,Capozzi2020PRD,Singh2020EPL,Barreiros2020,Kitabayashi2020PRD,Kitabayashi2017IJMPA,Kitabayashi2017IJMPA2,Kitabayashi2019IJMPA}, and textures under discrete symmetries, such as $A_n$ and $S_n$ \cite{Altarelli2010PMP}. The $\mu$-$\tau$ reflection symmetry has emerged as one of the most successful approaches for explaining the flavor neutrino mass matrix for Majorana neutrinos\cite{HarrisonPLB2002,XingRPP2023,XingRPP2016}.

In this study, we consider a unified mixing proposed in Ref.\cite{Kumar2013PRD}, which serves as the origin of several well-known neutrino mixing schemes: the tribimaximal mixing (TBM)\cite{Harrison2002PLB,Xing2002PLB}, baimaximal mixing (BM)\cite{Barger1998PLB, Altarelli1998JHEP, Mohapatra1999PRD}, golden ratio mixing of type I (GRMI)\cite{Kajiyama2007PRD}, golden ratio mixing of type II (GRMII)\cite{Rodejohann2009PLB}, hexagonal mixing (HM)\cite{Giunti2003NPBSUPPL,Xing2003JPG,Varzielas2011PRD,Albright2010EPJC}, and democratic mixing (DM) \cite{Fritzsch1996PLB}. We construct an approximate $\mu$-$\tau$ reflection symmetric flavor neutrino mass matrix based on the unified mixing scheme. Then, we investigate the effects of the $\mu$-$\tau$ reflection symmetry breaking in the mass matrix. Our analysis reveals that the predicted sum of neutrino masses derived from an approximate $\mu$-$\tau$ reflection symmetric flavor neutrino mass matrix based on the unified neutrino mixing with an inverted mass ordering, is excluded from DESI2024 and Supernova Ia luminosity distance data. This conclusion implies that TBM, BM, GRM, and HM, under an approximate $\mu$-$\tau$ reflection symmetry with an inverted mass ordering of neutrinos, are also excluded from observations.

The remainder of this paper is organized as follows. In Sect.\ref{Section:Lmm}, we review the unified mixing matrix proposed in Ref.\cite{Kumar2013PRD}. In Sect.\ref{Section:mutau}, we construct an approximate $\mu$-$\tau$ reflection symmetric flavor neutrino mass matrix based on the unified mixing scheme. In Sect. \ref{Section:NC}, we examine the experimental constraints on the approximate $\mu$-$\tau$ reflection symmetric flavor neutrino mass matrix. In Sec.\ref{Section:summary}, we present the summary.

\section{Unified neutrino mixing \label{Section:Lmm}}
\begin{table}
\tbl{Parameters $a$ and $b$ in the unified mixings.}
{\begin{tabular}{@{}c|cc|cc@{}} \toprule
&$U_{23}$&&$U_{13}$& \\
Mixing&$a$&$b$&$a$&$b$ \\
\hline
TBM&$2$&$1$&$1$&$1$\\
BM&$\sqrt{2}$&$1$&$\sqrt{2}$&$1$\\
GRM1&$\sqrt{3+\sqrt{5}}$&$1$&$\sqrt{3-\sqrt{5}}$&$1$\\
GRM2&$\sqrt{2+4/\sqrt{5}}$&$1$&$\sqrt{10-4\sqrt{5}}$&$1$\\
HM&$\sqrt{6}$&$1$&$\sqrt{2/3}$&$1$\\
DM&$\sqrt{3/2}$&$1/\sqrt{2}$&$\sqrt{3/2}$&$1/\sqrt{2}$\\
\hline
\hline
&$U^{13}$&&$U^{12}$& \\
Mixing&$a$&$b$&$a$&$b$ \\
\hline
TBM&$\sqrt{2}$&$\sqrt{2}$&$\sqrt{3}$&$\sqrt{2}$\\
BM&$\sqrt{2}$&$1$&$\sqrt{2}$&$1$\\
GRM1&$\sqrt{(5+\sqrt{5})/2}$&$\sqrt{(3+\sqrt{5})/2}$&$\sqrt{(5+\sqrt{5})/2}$&$\sqrt{(3+\sqrt{5})/2}$\\
GRM2&$\sqrt{2+2/\sqrt{5}}$&$1+\sqrt{5}/\sqrt{10-2/\sqrt{5}}$&$\sqrt{2+2/\sqrt{5}}$&$(1+\sqrt{5})/\sqrt{10-2/\sqrt{5}}$\\
HM&$1$&$\sqrt{3}$&$1$&$\sqrt{3}$\\
DM&$2$&$1$&$1$&$1$\\
\hline
\end{tabular} \label{tbl:Value_a_b} }
\end{table}
We briefly review the unified mixing, which serves as the origin of the TBM, BM, GRM, HM and DM, proposed in Ref. \cite{Kumar2013PRD}. This unified mixing scheme is realized by the following four type of matrices:
\begin{eqnarray}
U_{23}=
\left ( 
\begin{array}{ccc}
aN&N\sqrt{1+b^2}\cos{\theta}&N\sqrt{1+b^2}\sin{\theta}\\
bN&\frac{e^{i \phi}\sin{\theta}-abN\cos{\theta}}{\sqrt{1+b^2}}&-\frac{e^{i \phi}\cos{\theta}+abN\sin{\theta}}{\sqrt{1+b^2}}\\
N&-\frac{aN\cos{\theta}+be^{i \phi}\sin{\theta}}{\sqrt{1+b^2}}&\frac{be^{i \phi}\cos{\theta}-aN\sin{\theta}}{\sqrt{1+b^2}}\\
\end{array}
\right),
\label{Eq:U23}
\end{eqnarray}
\begin{eqnarray}
U_{13}=
\left ( 
\begin{array}{ccc}
N\sqrt{1+b^2}\cos{\theta}&aN&N\sqrt{1+b^2}\sin{\theta}\\
\frac{e^{i \phi}\sin{\theta}-abN\cos{\theta}}{\sqrt{1+b^2}}&bN&-\frac{e^{i \phi}\cos{\theta}+abN\sin{\theta}}{\sqrt{1+b^2}}\\
-\frac{aN\cos{\theta}+be^{i \phi}\sin{\theta}}{\sqrt{1+b^2}}&N&\frac{be^{i \phi}\cos{\theta}-aN\sin{\theta}}{\sqrt{1+b^2}}\\
\end{array}
\right),
\label{Eq:U13}
\end{eqnarray}
\begin{eqnarray}
U^{13}=
\left ( 
\begin{array}{ccc}
\frac{be^{i \phi}\cos{\theta}-aN\sin{\theta}}{\sqrt{1+b^2}}&-\frac{e^{i \phi}\cos{\theta}+abN\sin{\theta}}{\sqrt{1+b^2}}&N\sqrt{1+b^2}\sin{\theta}\\
N&bN&aN\\
-\frac{aN\cos{\theta}+be^{i \phi}\sin{\theta}}{\sqrt{1+b^2}}&\frac{e^{i \phi}\sin{\theta}-abN\cos{\theta}}{\sqrt{1+b^2}}&N\sqrt{1+b^2}\cos{\theta}\\
\end{array}
\right),
\label{Eq:U2row}
\end{eqnarray}
and
\begin{eqnarray}
U^{12}=
\left ( 
\begin{array}{ccc}
\frac{be^{i \phi}\cos{\theta}-aN\sin{\theta}}{\sqrt{1+b^2}}&-\frac{e^{i \phi}\cos{\theta}+abN\sin{\theta}}{\sqrt{1+b^2}}&N\sqrt{1+b^2}\sin{\theta}\\
-\frac{aN\cos{\theta}+be^{i \phi}\sin{\theta}}{\sqrt{1+b^2}}&\frac{e^{i \phi}\sin{\theta}-abN\cos{\theta}}{\sqrt{1+b^2}}&N\sqrt{1+b^2}\cos{\theta}\\
N&bN&aN\\
\end{array}
\right),
\label{Eq:U3row}
\end{eqnarray}
where $\theta$ denotes a rotation angle, $\phi$ denotes a phase parameter and $N=1/\sqrt{a^2+b^2+1}$. For example, the mixing matrix of TBM is obtained from $U_{23}$ with $a=2$, $b=1$, and $\theta=\phi=0$. Similarly, the mixing matrices of BM, GRM, HM and DM are obtained with the set of $(a,b)$ in Table. \ref{tbl:Value_a_b} and $\theta=\phi=0$.

In general, the sines and cosines of the three neutrino mixing angles $\theta_{ij}$ ($i,j$=1,2,3) are obtained as follows:
\begin{eqnarray}
s_{12}^2&=&\frac{|U_{e2}|^2}{1-|U_{e3}|^2},~~ s_{23}^2=\frac{|U_{\mu3}|^2}{1-|U_{e3}|^2},~~s_{13}^2=|U_{e3}|^2,\nonumber \\
c_{12}^2&=&\frac{|U_{e1}|^2}{1-|U_{e3}|^2},~~c_{23}^2=\frac{|U_{\tau3}|^2}{1-|U_{e3}|^2},
\label{Eq:mixing_angle_PMNS}
\end{eqnarray}
where $c_{ij}=\cos\theta_{ij}$, $s_{ij}=\sin\theta_{ij}$ ($i,j$=1,2,3). $U_{\alpha i}$ $(\alpha = e, \mu, \tau)$ is an element of mixing matrix $U$: \cite{Pontecorvo1957,Pontecorvo1958,Maki1962PTP,PDG}
\begin{eqnarray}
U  
&=&\left ( 
\begin{array}{ccc}
U_{e1}&U_{e2}&U_{e3}\\
U_{\mu1}&U_{\mu2}&U_{\mu3}\\
U_{\tau1}&U_{\tau2}&U_{\tau3}
\end{array}
\right)\nonumber\\
&=&
\left ( 
\begin{array}{ccc}
c_{12}c_{13} & s_{12}c_{13} & s_{13} e^{-i\delta} \\
- s_{12}c_{23} - c_{12}s_{23}s_{13} e^{i\delta} & c_{12}c_{23} - s_{12}s_{23}s_{13}e^{i\delta} & s_{23}c_{13} \\
s_{12}s_{23} - c_{12}c_{23}s_{13}e^{i\delta} & - c_{12}s_{23} - s_{12}c_{23}s_{13}e^{i\delta} & c_{23}c_{13} \\
\end{array}
\right),
\end{eqnarray}
where $\delta$ denotes the Dirac CP phase. To estimate the Dirac CP phase, we use the Jarlskog rephasing invariant\cite{Jarlskog},
\begin{eqnarray}
J=\rm{Im}(U_{e1}U_{e2}^{\ast}U_{\mu1}^{\ast}U_{\mu2})=s_{12}s_{23}s_{13}c_{12}c_{23}c_{13}^2\sin{\delta}.
\label{Eq:Jarlskog}
\end{eqnarray}
%

\begin{table}[t]
\tbl{The predicted mixing angles in the unified mixing.}
{\begin{tabular}{@{}cccc@{}} \toprule
Mixing matrix&$s_{12}^2$&$s_{23}^2$&$s_{13}^2$ \\
\hline
$U_{23}$ &$1-\frac{a^2}{a^2+(1+b^2)\cos^2{\theta}}$&$s_0^2+A\cos{\phi}$&$(1+b^2)N^2\sin^2{\theta}$ \\
\hline
$U_{13}$&$\frac{a^2}{a^2+(1+b^2)\cos^{\theta}}$&$s_0^2+A\cos{\phi}$&$(1+b^2)N^2\sin^2{\theta}$ \\
\hline
$U^{13}$&$s_0^2+A\cos{\phi}$&$\frac{a^2}{a^2+(1+b^2)\cos^{\theta}}$&$(1+b^2)N^2\sin^2{\theta}$ \\
\hline
$U^{12}$&$s_0^2+A\cos{\phi}$&$1-\frac{a^2}{a^2+(1+b^2)\cos^2{\theta}}$&$(1+b^2)N^2\sin^2{\theta}$ \\
\hline
\end{tabular} \label{tbl:mixing_angles}}
\end{table}

From Eq.(\ref{Eq:mixing_angle_PMNS}), the mixing angles of the unified mixing are predicted as shown in Table.\ref{tbl:mixing_angles} with 
\begin{eqnarray}
s_0^2=\frac{1}{1+b^2}\left(1-\frac{a^2(1+b^2)\sin^2{\theta}}{a^2+(1+b^2)\cos^2{\theta}}\right),
\end{eqnarray}
and
\begin{eqnarray}
A=\frac{ab\sin{2\theta}}{N(1+b^2)(a^2+(1+b^2)\cos^2{\theta})}.
\end{eqnarray}
%

\section{Approximate $\mu$-$\tau$ reflection symmetry \label{Section:mutau}}
In this section, we construct an approximate $\mu$-$\tau$ reflection symmetric flavor neutrino mass matrix based on the unified mixing.

The flavor neutrino mass matrix can be constructed via mixing matrix $U$ as follows:
\begin{eqnarray}
M&=&U{\rm diag}(m_1,m_2e^{2i\alpha},m_3e^{2i\beta})U^T,
\end{eqnarray}
where $\alpha$ and $\beta$ are the Majorana CP phases. The general texture of the exact $\mu$-$\tau$ reflection symmetric flavor neutrino mass matrix is  \cite{HarrisonPLB2002,XingRPP2023,XingRPP2016},
\begin{eqnarray}
M^{\mu-\tau}=
\left ( 
\begin{array}{ccc}
M_{ee}&M_{e\mu}&M_{e\mu}^{\ast}\\
M_{e\mu}&M_{\mu\mu}&M_{\mu\tau}\\
M_{e\mu}^{\ast}&M_{\mu\tau}&M_{\mu\mu}^{\ast}\\
\end{array}
\right).
\label{Eq:exact_mu_tau}
\end{eqnarray}

First, we omit the Majorana CP phases by setting $\alpha=\beta=0$. In this case, it is known that the condition $|U_{\mu i}|=|U_{\tau i}| (i=1,2,3)$ can sometime lead to an exact $\mu$-$\tau$ reflection symmetric flavor neutrino mass matrix. Under this requirement, we sometimes successfully get the exact $\mu$-$\tau$ reflection symmetric flavor neutrino mass matrix. However, in other cases, we derive a different type of mass matrix, such as a $\mu$-$\tau$ symmetric but non-reflection mass matrix. The outcome depends on the specific form of the mixing matrix $U$. It is worth exploring the application of the condition $|U_{\mu i}|=|U_{\tau i}|$ to elements such as $U_{23}, U_{13}, U^{13}$ and $U^{12}$ to derive unified mixing matrices capable of constructing an exact $\mu$-$\tau$ reflection symmetric flavor neutrino mass matrix.

The condition $|U_{\mu i}|=|U_{\tau i}|$ can be satisfied by setting $b=1$ and $\phi=270^\circ$ in the mixing matrices $U_{23}$ and $U_{13}$. The unified mixing matrices $U_{23}$ and $U_{13}$, which satisfy the condition $|U_{\mu i}|=|U_{\tau i}|$, are obtained as follows:
\begin{eqnarray}
\tilde{U}_{23}=
\left ( 
\begin{array}{ccc}
aN&\sqrt{2}N\cos{\theta}&\sqrt{2}N\sin{\theta}\\
N&-\frac{aN\cos{\theta}+i\sin{\theta}}{\sqrt{2}}&\frac{-aN\sin{\theta}+i\cos{\theta}}{\sqrt{2}}\\
N&\frac{-aN\cos{\theta}+i\sin{\theta}}{\sqrt{2}}&-\frac{aN\sin{\theta}+i\cos{\theta}}{\sqrt{2}}\\
\end{array}
\right),
\label{Eq:U23_mutau}
\end{eqnarray}
and
\begin{eqnarray}
\tilde{U}_{13}=
\left ( 
\begin{array}{ccc}
\sqrt{2}N\cos{\theta}&aN&\sqrt{2}N\sin{\theta}\\
-\frac{aN\cos{\theta}+i\sin{\theta}}{\sqrt{2}}&N&\frac{-aN\sin{\theta}+i\cos{\theta}}{\sqrt{2}}\\
\frac{-aN\cos{\theta}+i\sin{\theta}}{\sqrt{2}}&N&-\frac{aN\sin{\theta}+i\cos{\theta}}{\sqrt{2}}\\
\end{array}
\right),
\label{Eq:U23_mutau}
\end{eqnarray}
respectively. On the other hand, for the unified mixing matrices $U^{13}$ and $U^{12}$, the condition $|U_{\mu i}|=|U_{\tau i}|$ is satisfied when $\theta=0$. The unified mixing matrices $U^{13}$ and $U^{12}$, which satisfy the condition $|U_{\mu i}|=|U_{\tau i}|$, are obtained as follows:
\begin{eqnarray}
\tilde{U}^{13}=
\left ( 
\begin{array}{ccc}
\frac{be^{i\phi}}{\sqrt{b^2+1}} &-\frac{e^{i\phi}}{\sqrt{b^2+1}}&0\\
\frac{\sqrt{2}}{2\sqrt{b^2+1}} &\frac{\sqrt{2}b}{2\sqrt{b^2+1}}&\frac{\sqrt{2}}{2}\\
-\frac{\sqrt{2}}{2\sqrt{b^2+1}}  &-\frac{\sqrt{2}b}{2\sqrt{b^2+1}}&\frac{\sqrt{2}}{2}\\
\end{array}
\right),
\end{eqnarray}
and
\begin{eqnarray}
\tilde{U}^{12}=
\left ( 
\begin{array}{ccc}
\frac{be^{i\phi}}{\sqrt{b^2+1}}  &-\frac{e^{i\phi}}{\sqrt{b^2+1}}&0\\
-\frac{\sqrt{2}}{2\sqrt{b^2+1}}&-\frac{\sqrt{2}b}{2\sqrt{b^2+1}}&\frac{\sqrt{2}}{2}\\
\frac{\sqrt{2}}{2\sqrt{b^2+1}}&\frac{\sqrt{2}b}{2\sqrt{b^2+1}}&\frac{\sqrt{2}}{2}\\
\end{array}
\right),
\end{eqnarray}
respectively.

To investigate whether an exact $\mu$-$\tau$ reflection symmetric flavor neutrino mass matrix can be derived from these matrices, $\tilde{U}_{23}$, $\tilde{U}_{13}$, $\tilde{U}^{13}$ and $\tilde{U}^{12}$, we construct the flavor neutrino mass matrices, $\tilde{M}_{23}$, $\tilde{M}_{13}$, $\tilde{M}^{13}$, and $\tilde{M}^{12}$ as follows:
\begin{eqnarray}
\tilde{M}_{23} &=& \tilde{U}_{23}{\rm diag}(m_1,m_2,m_3)\tilde{U}_{23}^T, \nonumber \\
\tilde{M}_{13} &=& \tilde{U}_{13}{\rm diag}(m_1,m_2,m_3)\tilde{U}_{13}^T, \nonumber \\
\tilde{M}^{13} &=& \tilde{U}^{13}{\rm diag}(m_1,m_2,m_3)(\tilde{U}^{13})^T, \nonumber \\
\tilde{M}^{12} &=& \tilde{U}^{22}{\rm diag}(m_1,m_2,m_3)(\tilde{U}^{12})^T.
\end{eqnarray}
We have found that the textures of $\tilde{M}_{23}$ and $\tilde{M}_{13}$ are same as the texture of exact $\mu$-$\tau$ reflection symmetric flavor neutrino mass matrix which is shown in Eq.(\ref{Eq:exact_mu_tau}):
\begin{eqnarray}
\tilde{M}_{23} =\tilde{M}_{13}=
\left ( 
\begin{array}{ccc}
M_{ee}&M_{e\mu}&M_{e\mu}^{\ast}\\
M_{e\mu}&M_{\mu\mu}&M_{\mu\tau}\\
M_{e\mu}^{\ast}&M_{\mu\tau}&M_{\mu\mu}^{\ast}\\
\end{array}
\right)
=M^{\mu-\tau}.
\end{eqnarray}
On the contrary, $\tilde{M}^{13}$ and $\tilde{M}^{23}$ have the following form  
\begin{eqnarray}
\tilde{M}^{13}=\tilde{M}^{12}=
\left ( 
\begin{array}{ccc}
M_{ee} & M_{e\mu} &-M_{e\mu} \\
M_{e\mu} & M_{\mu\mu}&M_{\mu\tau} \\
-M_{e\mu}& M_{\mu\tau} &M_{\mu\mu} \\
\end{array}
\right)
\neq M^{\mu-\tau},
\end{eqnarray}
and these mass matrices are not under exact $\mu$-$\tau$ reflection symmetry. Therefore, we omit the mixing matrices $\tilde{U}^{13}$ and $\tilde{U}^{12}$, in addition to the mass matrices $\tilde{M}^{13}$ and $\tilde{M}^{12}$, in the remainder of our discussion.

From this point onward, we include arbitrary Majorana phases $\alpha$ and $\beta$ in the mass matrix. We expect that non-zero Majorana CP phases cause a violation of the $\mu$-$\tau$ reflection symmetry \cite{TostadoNPB2023,TostadoarXiv2024,XingRPP2016, TostadoNPB2021,ZhaoJHEP2017}. The flavor neutrino mass matrix derived from $\tilde{U}_{23}$, incorporating arbitrary Majorana CP phases is obtained as follows:
\begin{eqnarray}
M_{23}&=&\tilde{U}_{23}{\rm diag}(m_1,m_2e^{2i\alpha},m_3e^{2i\beta})\tilde{U}_{23}^T  \nonumber \\
&=&M_{23}^{\mu-\tau}+\Delta M_{23},
\label{Eq:M23}
\end{eqnarray}
where 
\begin{eqnarray}
\hspace{-0.3cm}
\scriptsize 
M_{23}^{\mu-\tau}=\left(
\begin{array}{ccc}
M_{ee}&M_{e\mu}&M_{e\mu}^{\ast}\\
M_{e\mu}&M_{\mu\mu}&M_{ee}-\left(a-\frac{1}{a}\right)M_{e\mu}+\frac{1}{a}M_{e\mu}^{\ast}-M_{\mu\mu}\\
M_{e\mu}^{\ast}&M_{ee}-\left(a-\frac{1}{a}\right)M_{e\mu}+\frac{1}{a}M_{e\mu}^{\ast}-M_{\mu\mu}&M_{\mu\mu}^{\ast}\\
\end{array}
\right), 
\end{eqnarray}
and
\begin{eqnarray}
\Delta M_{23}&=&\left ( 
\begin{array}{ccc}
0&0&M_{e\tau}-M_{e\mu}^{\ast}\\
0&0&\frac{1}{a}(M_{e\tau}-M_{e\mu}^{\ast})\\
M_{e\tau}-M_{e\mu}^{\ast}&\frac{1}{a}(M_{e\tau}-M_{e\mu}^{\ast})&a(M_{e\mu}-M_{e\tau})+M_{\mu\mu}-M_{\mu\mu}^{\ast}\\
\end{array}
\right)\nonumber\\
&=&\left ( 
\begin{array}{ccc}
0&0&\epsilon_{23}^1\\
0&0&\frac{1}{a}\epsilon_{23}^1\\
\epsilon_{23}^1&\frac{1}{a}\epsilon_{23}^1&\epsilon_{23}^2\\
\end{array}
\right).
\label{Eq:delta_23}
\end{eqnarray}
Here, $M_{23}^{\mu-\tau}$ satisfies the exact $\mu$-$\tau$ reflection symmetry, while $\Delta M_{23}$ is a matrix representing the deviation from the exact $\mu$-$\tau$ reflection symmetry. The explicit expressions of matrix elements and symmetry breaking term $\epsilon_{23}^i$ $(i=1,2)$ are the following:
\begin{eqnarray}
M_{ee}&=&\frac{a^{2} m_{1} + 2 m_{2} e^{2 i \alpha} \cos^{2}{\theta} + 2 m_{3} e^{2 i \beta} \sin^{2}{\theta}}{a^{2} + 2}, \nonumber \\
M_{e\mu}&=&\frac{1}{2(a^2+2)}\times \biggl\{ 2am_1-m_{2} \left(a \cos{2 \theta} + a + i \sqrt{a^{2} + 2} \sin{2 \theta}\right) e^{2 i \alpha}\nonumber\\
&&-m_{3} \left(- a \cos{2 \theta} + a - i \sqrt{a^{2} + 2} \sin{2 \theta}\right) e^{2 i \beta}\biggr\}\nonumber\\
M_{\mu\mu}&=&\frac{1}{2(a^2+2)}\times \biggl\{2 m_{1} + m_{2} \left(a \cos{\theta} + i \sqrt{a^{2} + 2} \sin{\theta}\right)^{2} e^{2 i \alpha} \nonumber\\
&&+ m_{3} \left(a \sin{\theta} - i \sqrt{a^{2} + 2} \cos{\theta}\right)^{2} e^{2 i \beta}\biggr\}\nonumber\\
M_{e\tau}&=&\frac{1}{2(a^2+2)}\times \biggl\{ 2am_1-m_{2} \left(a \cos{2 \theta} + a - i \sqrt{a^{2} + 2} \sin{2 \theta}\right) e^{2 i \alpha}\nonumber\\
&&-m_{3} \left(- a \cos{2 \theta} + a + i \sqrt{a^{2} + 2} \sin{2 \theta}\right) e^{2 i \beta}\biggr\}
\end{eqnarray}
and
\begin{eqnarray}
\epsilon_{23}^1&=&m_2\left(a+a\cos{2\theta}-i\sqrt{a^2+2}\sin{2\theta}\right)i\sin{2\alpha}\nonumber\\
&&+m_3\left(a-a\cos{2\theta}+i\sqrt{a^2+2}\sin{2\theta}\right)i\sin{2\beta},\nonumber \\
\epsilon_{23}^2&=&aC_1+C_2,
\end{eqnarray}
respectively, where 
\begin{eqnarray}
C_1&=&\frac{i \left(- m_{2} e^{2 i \alpha} + m_{3} e^{2 i \beta}\right) \sin{2 \theta}}{\sqrt{a^{2} + 2}},\nonumber \\
C_2&=&m_{2} \left(- \frac{\left(a \cos{\theta} - i \sqrt{a^{2} + 2} \sin{\theta}\right)^{2} e^{- 2 i \alpha}}{2 a^{2} + 4} + \frac{\left(a \cos{\theta} + i \sqrt{a^{2} + 2} \sin{\theta}\right)^{2} e^{2 i \alpha}}{2 a^{2} + 4}\right) \nonumber \\
&&+ m_{3} \left(\frac{\left(a \sin{\theta} - i \sqrt{a^{2} + 2} \cos{\theta }\right)^{2} e^{2 i \beta}}{2 a^{2} + 4} - \frac{\left(a \sin{\theta} + i \sqrt{a^{2} + 2} \cos{\theta }\right)^{2} e^{- 2 i \beta}}{2 a^{2} + 4}\right).\nonumber \\
\end{eqnarray}
Note that if we set $\alpha=\beta=0^\circ, 90^\circ, 180^\circ, 270^\circ$, and $360^\circ$, $\epsilon_{23}^1=\epsilon_{23}^2=0$ as well as $\Delta M_{23}=0$ are obtained and the exact $\mu$-$\tau$ reflection symmetry is recovered.

On the other hand, the flavor neutrino mass matrix via $\tilde{U}_{13}$ with arbitrary Majorana CP phases is obtained as:
\begin{eqnarray}
M_{13}&=&\tilde{U}_{13}{\rm diag}(m_1,m_2e^{2i\alpha},m_3e^{2i\beta})\tilde{U}_{13}^T \nonumber \\
&=&M_{13}^{\mu-\tau}+\Delta M_{13},
\end{eqnarray}
where 
\begin{eqnarray}
\hspace{-0.3cm}
\scriptsize 
M_{13}^{\mu-\tau}=\left(
\begin{array}{ccc}
M_{ee}&M_{e\mu}&M_{e\mu}^{\ast}\\
M_{e\mu}&M_{\mu\mu}&M_{ee}-\left(a-\frac{1}{a}\right)M_{e\mu}+\frac{1}{a}M_{e\mu}^{\ast}-M_{\mu\mu}\\
M_{e\mu}^{\ast}&M_{ee}-\left(a-\frac{1}{a}\right)M_{e\mu}+\frac{1}{a}M_{e\mu}^{\ast}-M_{\mu\mu}&M_{\mu\mu}^{\ast}\\
\end{array}
\right), 
\end{eqnarray}
and
\begin{eqnarray}
\Delta M_{13}&=&\left ( 
\begin{array}{ccc}
0&0&M_{e\tau}-M_{e\mu}^{\ast}\\
0&0&\frac{1}{a}(M_{e\tau}-M_{e\mu}^{\ast})\\
M_{e\tau}-M_{e\mu}^{\ast}&\frac{1}{a}(M_{e\tau}-M_{e\mu}^{\ast})&a(M_{e\mu}-M_{e\tau})+M_{\mu\mu}-M_{\mu\mu}^{\ast}\\
\end{array}
\right)\nonumber\\
&=&\left ( 
\begin{array}{ccc}
0&0&\epsilon_{13}^1\\
0&0&\frac{1}{a}\epsilon_{13}^1\\
\epsilon_{13}^1&\frac{1}{a}\epsilon_{13}^1&\epsilon_{13}^2\\
\end{array}
\right).
\label{Eq:delta_13}
\end{eqnarray}
The explicit expressions of matrix elements and symmetry breaking term $\epsilon_{13}^i$ $(i=1,2)$ are the following:
\begin{eqnarray}
M_{ee}&=&N^{2} \left(2 m_{1} \cos^{2}{\theta} +a^{2} m_{2} e^{2 i \alpha}+ 2 m_{3} e^{2 i \beta} \sin^{2}{\theta}\right),\nonumber \\
M_{e\mu}&=&-N\biggl(\frac{m_{1} \left(aN \cos{2 \theta} + aN  + i \sin{2 \theta}\right)}{2}-aN m_{2} e^{2 i \alpha}\nonumber \\
&&+\frac{m_{3} \left(- aN\cos{2 \theta} + aN - i \sin{2 \theta}\right) e^{2 i \beta}}{2}\biggr),\nonumber \\
M_{\mu\mu}&=& \frac{m_{1} \left(aN \cos{\theta} + i \sin{\theta}\right)^{2}}{2} +N^{2} m_{2} e^{2 i \alpha}+ \frac{m_{3} \left(aN\sin{\theta} - i \cos{\theta }\right)^{2} e^{2 i \beta}}{2}, \nonumber \\
M_{e\tau}&=&- N\biggl(\frac{m_{1} \left(aN\cos{2 \theta} + aN - i \sin{2 \theta}\right)}{2}-aN m_{2} e^{2 i \alpha}\nonumber\\
&&+\frac{m_{3} \left(- aN \cos{2 \theta} + aN+ i \sin{2 \theta}\right) e^{2 i \beta}}{2}\biggr),
\end{eqnarray}
and
\begin{eqnarray}
\epsilon_{13}^1&=&aN^2m_2 i\sin(2\alpha)+m_3\left(aN^2\cos(2\theta)-aN^2-iN\sin(2\theta)\right)i\sin(2\beta),\nonumber \\
\epsilon_{13}^2&=&aC_1+C_2,
\end{eqnarray}
respectively, where
\begin{eqnarray}
C_1&=&N \left(- m_{1} + m_{3} e^{2 i \beta}\right) i\sin{2 \theta},\nonumber \\
C_2&=&N a m_{1} i\sin{2 \theta} +2N^2 m_{2}  i\sin{2 \alpha}\nonumber \\
&& +m_{3} \left(\frac{\left(aN\sin{\theta} - i \cos{\theta }\right)^{2} e^{2 i \beta}}{2}-\frac{\left(aN\sin{\theta} +i \cos{\theta }\right)^{2} e^{-2 i \beta}}{2}\right).
\end{eqnarray}
Similarly to the case of $M_{23}$, if we set $\alpha=\beta=0^\circ, 90^\circ, 180^\circ, 270^\circ$, and $360^\circ$, $\epsilon_{13}^1=\epsilon_{13}^2=0$ as well as $\Delta M_{13}=0$ are obtained and the exact $\mu$-$\tau$ reflection symmetry is recovered.

We define the dimensionless parameters to characterize the violations of the $\mu$-$\tau$ reflection symmetry:
\begin{eqnarray}
\epsilon_{23} = \left| \frac{\epsilon_{23}^1}{(M_{23})_{e\mu}^{\ast}} \right| +  \left| \frac{\epsilon_{23}^2}{(M_{23})_{\mu\mu}^{\ast}} \right|,
\quad 
\epsilon_{13} = \left| \frac{\epsilon_{13}^1}{(M_{13})_{e\mu}^{\ast}} \right| +  \left| \frac{\epsilon_{13}^2}{(M_{13})_{\mu\mu}^{\ast}} \right|.
\end{eqnarray}
If $0<\epsilon_{23}\leq 0.1$ and $0<\epsilon_{13}\leq 0.1$, we deem that the flavor neutrino mass matrix $M_{23}$ and $M_{13}$ are satisfied with an approximate $\mu$-$\tau$ reflection symmetry, respectively.
\section{Experimental constraints \label{Section:NC}}
A global analysis of the current data provides the following best-fit values of the neutrino mixing angles in the case of NO, where $m_1,m_2,m_3$ represent the neutrino mass eigenvalues \cite{NuFit}:
\begin{eqnarray} 
s_{12}^2&=& 0.308^{+0.012}_{-0.011} \quad (0.275 \sim 0.345), \nonumber \\
s_{23}^2&=& 0.451^{+0.017}_{-0.013} \quad (0.435 \sim 0.585), \nonumber \\
s_{13}^2&=& 0.02215^{+0.00056}_{-0.00058} \quad (0.02030 \sim 0.02388), \nonumber \\
\delta/^\circ &=& 212^{+26}_{-19} \quad (124 \sim 364), 
\label{Eq:neutrino_observation_NO}
\end{eqnarray}
where $\pm$ denotes the $1 \sigma$ region and the parentheses denote the $3 \sigma$ region. For the IO, $m_3 < m_1<m_2$, we obtain:
\begin{eqnarray} 
s_{12}^2&=&  0.308^{+0.012}_{-0.011} \quad (0.275 \sim 0.345), \nonumber \\
s_{23}^2 &=& 0.550^{+0.012}_{-0.015} \quad (0.440 \sim 0.584), \nonumber \\
s_{13}^2&=& 0.02231^{+0.00056}_{-0.00056} \quad (0.02060 \sim 0.02409), \nonumber \\
\delta/^\circ &=& 274^{+22}_{-25} \quad (201  \sim 335).
\label{Eq:neutrino_observation_IO}
\end{eqnarray}
Moreover, we have the following constraints:
\begin{eqnarray} 
\sum m_i < 0.12 - 0.69 ~{\rm eV},
\end{eqnarray}
from the cosmological observation of the cosmic microwave background radiation\cite{Planck2018, Capozzi2020PRD,Giusarma2016PRD,Vagnozzi2017PRD,Giusarma2018PRD}.

We require that the square mass differences $\Delta m_{ij}^{2}$, mixing angles $\theta_{ij}$, and the Dirac CP-violating phase $\delta$ derived from $M_{23}$ and $M_{13}$ fall within the observed $3\sigma$ region. Furthermore, the lightest neutrino mass ,$m_{\rm{lightest}}$ ($m_{\rm{lightest}}=m_1$ for NO and $m_{\rm{lightest}}=m_3$ for IO), is to lie within the range $0-0.12$ eV. We also impose the constraint $\sum m_i < 0.12 ~{\rm eV}$. In our analysis, we vary $\alpha, \beta$ over their full range, e.g., $0^\circ  - 360^\circ$. The value of parameter $a$ in the mixing matrix is taken within the range $0.5 \le a \le 2.5$.

\begin{figure}[t]
\hspace{-1.5cm}
\begin{minipage}{0.5\textwidth}
\includegraphics[keepaspectratio, scale=0.7]{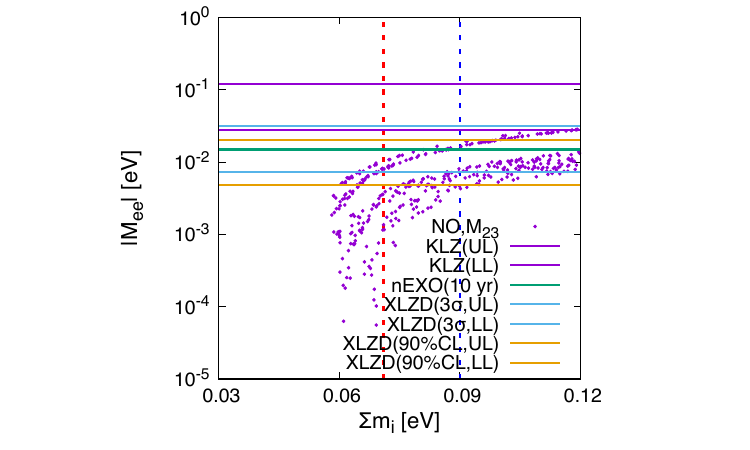}
\end{minipage}
\hfill
\begin{minipage}{0.6\textwidth}
\centering
\includegraphics[keepaspectratio, scale=0.7]{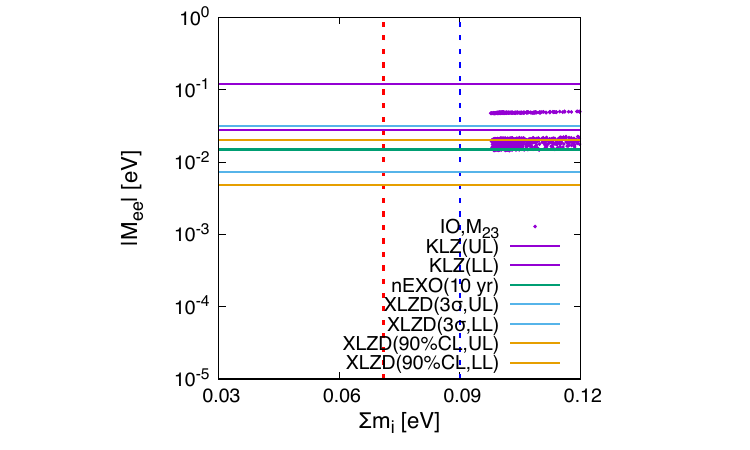}
\end{minipage}
\vspace*{8pt}
\caption{Relationship between the effective neutrino mass of the neutrinoless double $\beta$ decay $|M_{ee}|$ and the sum of the neutrino masses $\sum m_i$ in $M_{23}$ for the NO (left panel) and IO (right panel), where the $\epsilon_{23} \le 0.1$ condition needs to be met.\protect\label{fig:Mee_M23}}
\end{figure}

\begin{figure}[t]
\hspace{-1.5cm}
\begin{minipage}{0.5\textwidth}
\includegraphics[keepaspectratio, scale=0.7]{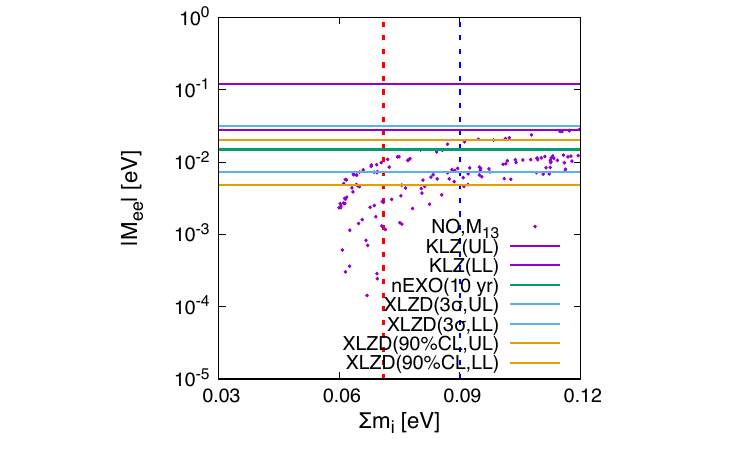}
\end{minipage}
\hfill
\begin{minipage}{0.6\textwidth}
\centering
\includegraphics[keepaspectratio, scale=0.7]{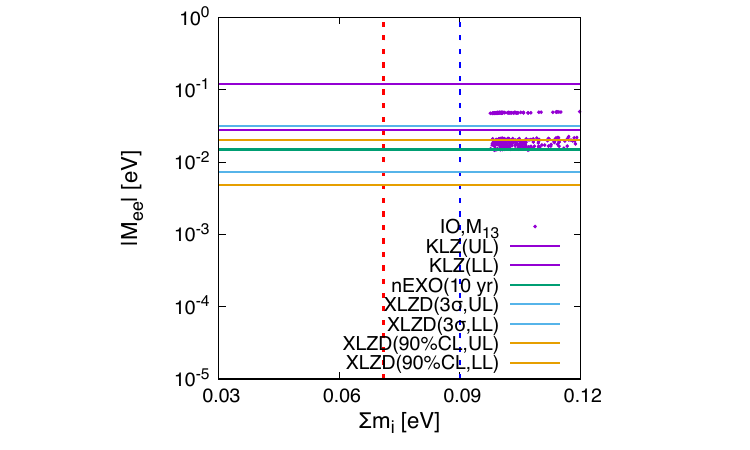}
\end{minipage}
\vspace*{8pt}
\caption{Same as Fig. \ref{fig:Mee_M23} but in $M_{13}$ with the condition of $\epsilon_{13} \le 0.1$.\protect \label{fig:Mee_M13}}
\end{figure}

Figure.\ref{fig:Mee_M23} illustrates the relationship between the effective neutrino mass for neutrinoless double $\beta$ decay $|M_{ee}|$ and the sum of the neutrino masses $\sum m_i$ in $M_{23}$ for the NO and IO cases for $\epsilon<0.1$. Figure.\ref{fig:Mee_M13} shows the same relationship as Fig.\ref{fig:Mee_M23}, but for $M_{13}$ under the condition $\epsilon_{13} \le 0.1$. In both figures, the purple, green, blue, and yellow horizontal lines represent the sensitivity of $|M_{ee}|$ in the KamLAND-Zen (KLZ) \cite{KamLand2024}, nEXO\cite{nEXO2017}, XLZD (3$\sigma$), and XLZD (90\%C. L.)\cite{XLZD2024}, respectively. The red and blue dotted vertical lines indicate the upper limits on the sum of the neutrino masses from DESI2024\cite{DESI2024} and Supernovae Ia luminosity distance data \cite{ValentinPRD2021}, respectively. From Fig.\ref{fig:Mee_M23} and Fig.\ref{fig:Mee_M13}, we observe that the predicted $\sum m_i$ derived from the approximate $\mu$-$\tau$ reflection symmetric neutrino mass matrices $M_{23}$ and $M_{13}$ for IO are excluded from the DESI2024 and Supernova Ia luminosity distance data. In addition, the $|M_{ee}|$ derived in the IO case will be subject to more severe restrictions from the experiments than in the NO case.

It is worth noting that all specific values of parameter $a$ for $U_{23}$ and $U_{13}$ listed in Table \ref{tbl:Value_a_b}, such as $a=2$ in $U_{23}$ for TBM, are within the range of $0.5 \le a \le 2.5$. Therefore, the conclusions reached so far also hold for TBM, BM, GRM1, GRM2, and HM under an approximate $\mu$-$\tau$ reflection symmetry (Note that we take $b=1$. Since DM needs $b \neq 1$, it is excluded). Thus, TBM, BM, GRMI, GRMII, and HM under an approximate $\mu$-$\tau$ reflection symmetry for IO are excluded from observational data.

\section{Summary\label{Section:summary}}
A unified mixing scheme, which serves as the origin of TBM, BM, GRM, HM, and DM was proposed in Ref. \cite{Kumar2013PRD}. In this study, we constructed an approximate $\mu$-$\tau$ reflection symmetric flavor neutrino mass matrix based on this unified mixing scheme. The effects of the $\mu$-$\tau$ reflection symmetry breaking in the mass matrix were analyzed. It was found that the predicted sum of neutrino masses derived from the approximate $\mu$-$\tau$ reflection symmetric neutrino mass matrix with inverted mass ordering, is excluded from DESI2024 and Supernova Ia luminosity distance data. This result means that TBM, BM, GRMI, GRMII, and HM are excluded from the observational data under an approximate $\mu$-$\tau$ reflection symmetry with inverted mass order of the neutrinos.

\end{document}